\documentclass[10pt,letterpaper]{article}
\usepackage[top=0.85in,left=1.75in,footskip=0.75in]{geometry}

\usepackage{amsmath,amssymb}
\usepackage{float}

\usepackage{changepage}

\usepackage[utf8x]{inputenc}

\usepackage{textcomp,marvosym}

\usepackage{cite}

\usepackage{nameref,hyperref}

\usepackage[right]{lineno}

\usepackage{microtype}
\DisableLigatures[f]{encoding = *, family = * }

\usepackage[table]{xcolor}

\usepackage{array}

\newcolumntype{+}{!{\vrule width 2pt}}

\newlength\savedwidth

\raggedright
\usepackage[aboveskip=1pt,labelfont=bf,labelsep=period,justification=raggedright,singlelinecheck=off]{caption}

\makeatletter
\renewcommand{\@biblabel}[1]{\quad#1.}
\makeatother

\usepackage{lastpage,fancyhdr,graphicx}
\usepackage{epstopdf}
\fancyheadoffset[L]{2.25in}
\fancyfootoffset[L]{2.25in}
\newcommand*{\myfont}{\fontfamily{Courier}\selectfont}

\begin{document}
\vspace*{0.2in}

\begin{flushleft}
{\Large
\textbf\newline{The Human Geography of Twitter} % Please use "sentence case" for title and headings (capitalize only the first word in a title (or heading), the first word in a subtitle (or subheading), and any proper nouns).
}
\newline
\\
Rudy Arthur\textsuperscript{1*},
Hywel T.P. Williams\textsuperscript{1},
\\
\bigskip
\textbf{1} Department of Computer Science, CEMPS, University of Exeter, EX4 4QE, UK
\\
\bigskip

% Insert additional author notes using the symbols described below. Insert symbol callouts after author names as necessary.
% 
% Remove or comment out the author notes below if they aren't used.
%
% Primary Equal Contribution Note
%\Yinyang These authors contributed equally to this work.

% Additional Equal Contribution Note
% Also use this double-dagger symbol for special authorship notes, such as senior authorship.
%\ddag Conceived project and wrote paper.

%\textcurrency Performed analysis % change symbol to "\textcurrency a" if more than one current address note
% \textcurrency b Insert second current address 
% \textcurrency c Insert third current address


% Group/Consortium Author Note
%\textpilcrow Membership list can be found in the Acknowledgments section.

% Use the asterisk to denote corresponding authorship and provide email address in note below.
* rudy.d.arthur@gmail.com

\end{flushleft}
\section*{Abstract}
Given the centrality of regions in social movements, politics and public administration we aim to quantitatively study inter- and intra-regional communication for the first time. This work uses social media posts to first identify contiguous geographical regions with a shared social identity and then investigate patterns of communication within and between them. Our case study uses over 150 days of located Twitter data from England and Wales. In contrast to other approaches, (e.g. phone call data records or online friendship networks) we have the message contents as well as the social connection. This allows us to investigate not only the volume of communication but also the sentiment and vocabulary. We find that the South-East and North-West regions are the most talked about; regions tend to be more positive about themselves than about others; people talk politics much more between regions than within. This methodology gives researchers a powerful tool to study identity and interaction within and between social-geographic regions.

% Please keep the Author Summary between 150 and 200 words
% Use first person. PLOS ONE authors please skip this step. 
% Author Summary not valid for PLOS ONE submissions.   
%\section*{Author summary}
%We use social media data to study communication within and between regions of England and Wales. Our methodology allows us to detect emergent, spatially contiguous regions from the social network and then study the vocabulary, volume and sentiment of their communication. This study describes a new methodology to quantitatively address questions of regional identity, rivalry and hierarchy for the first time.

%\linenumbers

% Use "Eq" instead of "Equation" for equation citations.
\section*{Introduction}
Studies of human social interaction using phone call data and online social networks \cite{Stephens2014,Takhteyev2012,Lengyel2015,Ratti2010,Blondel2015,Yin2017,Sobolevsky2013} have found that, contrary to some expectations \cite{Cairncross1997}, geography is alive and well. Despite digital technology decoupling distance and difficulty of communication, spatial proximity remains one of the key factors in determining who communicates with whom. Regions determined from records of telephone communication closely reflect traditional regional and local identities \cite{Ratti2010}. This has been confirmed for numerous countries \cite{Sobolevsky2013} and for various different forms of electronic interaction \cite{Lengyel2015}.

The notion of a region therefore has much more than a purely bureaucratic meaning.  Discussions of regional identity pervade social theory \cite{Paasi2003,Paasi2009}, and many stereotypes, sporting rivalries and political differences occur at the regional level. In this paper we will study the regions of England and Wales, which is of particular relevance at a time when British national identity is being challenged by Brexit, regional devolution, and the economic disparity between North and South. However given that national and international policies are often implemented at the regional level (e.g. the European Union (EU) cohesion policy \footnote{http://ec.europa.eu/regional\_policy/en/faq Accessed June 2018}), questions of regional identity have wider geo-political relevance. Given that geographic regions are so fundamental, our question is: how can we quantitatively study ideas like `regional identity', `regional rivalry' or the `cultural dominance' of regions?

We begin with the observation that online social networks tend to have similar properties to offline, spatially embedded, social networks. In fact spatial structure in communication networks is robust enough to have instrumental value. Much recent research, especially on social media, has focused on exploiting the strong spatial correlations that exist in friendship networks to infer the locations of users (e.g. \cite{Backstrom2010,Sadilek2012,jurgens2013}). Other work linking social networks to geography has, for example, attempted to determine the amount of commerce in a given area \cite{Porta2009} or the location of a city's `heart' \cite{Louail2014}. This field of network geography has mostly focused on the network's topology and how this influences interaction and accessibility \cite{batty2005}. It is our aim to move beyond this by studying a social network where the links carry much richer metadata.  

We analyse a social network of interactions on Twitter. This social network is constructed from `mention' interactions, in which one user explicitly mentions or replies to one or more other users, and thus it has a few interesting properties. Firstly, connections are intrinsically directional. Alice can mention Bob on Twitter without Bob's permission, and Bob does not have to reciprocate. This allows for asymmetries in communication, so (at the network level) some regions can be the target of more mentions than others. Secondly, and crucially, unlike either phone call networks (where the call content is unknown) or friend/follower networks (which do not imply communication) here we have both the directed link between users {\it and} the content of the message. 

The plan of the paper is as follows. We first demonstrate that user communities, identified algorithmically from Twitter mentions, are geographically contiguous and loosely correspond to our expectations, based on administrative boundaries and `folk' conceptions of British regions. This approach makes no {\it a priori} assumptions about the number, location or boundaries of different regions, and is independent of administrative demarcations that may or may not reflect real regional identities. Next we use these emergent regions as the subjects of a comparative study of intra- and inter-region communication. We will compare the vocabulary and topics used by members of a region when speaking to each other compared with those used to speak to `outsiders'. We will then look at the volume and sentiment of messages sent within and between regions. Thus we can ask questions like ``What does the South-West say to North Yorkshire?'' and answer them in a concrete way e.g. ``They talk about sport, and the sentiment of the communication is slightly more negative than average''.

\section*{Materials and methods}
Our dataset of tweets from all of England and Wales (defined by a bounding box with lower left longitude and latitude (-5.8,49.9) and upper right (-1.2,55.9)) was obtained from four separate collections, one for the South-West (which was ongoing from previous work \cite{us}) and the other three chosen to sample an area containing $\sim15$ million people each. This is to avoid any potential to hit the rate limit for the Twitter streaming API (1\% of all tweets globally). This collection lasted from 01/10/2017 to 22/03/2018. 

All of our tweets have geographical information attached as GPS co-ordinates or `place-tags'. Previous work \cite{us} has shown that GPS tagged posts are predominantly shares from other social media platforms or automated accounts, while place-tagged tweets represent direct human interactions on Twitter. Thus we exclusively use place tags for location and discard GPS-tagged tweets. We locate users by assigning them to grid tiles proportionally to the frequency of their tweeting within the tile. For example, we can have 0.5 of a user in tile 1 and 0.5 in tile 2 if that user tweets equally often from 1 and 2. This is preferred to using the user location field which is often blank, doesn't contain location information or is too vague (e.g. `England'). We end up with 4513957 useful tweets authored by users in England and Wales and which mention users in England or Wales (excluding self mentions). All of our analyses are performed with this set.

\section*{Identifying Regions with Tweets}

In our set of $\sim$4.5 million tweets there are $\sim375,000$ unique users who mention another user in the target area. The mention network is constructed by treating each grid tile as a node and then adding an edge, $e_{ab}$, between every pair of tiles $a$ and $b$ when a user in tile $a$ mentions a user in tile $b$. Edges are directed and have weight equal to the number of mentions sent from tile $a$ to tile $b$. Self-edges (i.e. $a=b$) are allowed.

The Louvain method \cite{Blondel2009} is used to find communities within the resulting network; this method of community detection is robust, fast and automatically determines the best (modularity maximising) number of communities. However, this method is intended to work on undirected graphs. To turn the directed mention network into an undirected graph we set the edge weight between every pair of tiles as the total number of tweets sent in either direction ($e_{ab}+e_{ba}$), ignoring self-edges. We run the Louvain algorithm with 100 random restarts (to sample multiple local maxima) and choose the community partition with highest modularity from the set of 100 outcomes. 

\begin{figure}[H]
\centering
\includegraphics[height=0.5\textheight, width=0.7\textwidth]{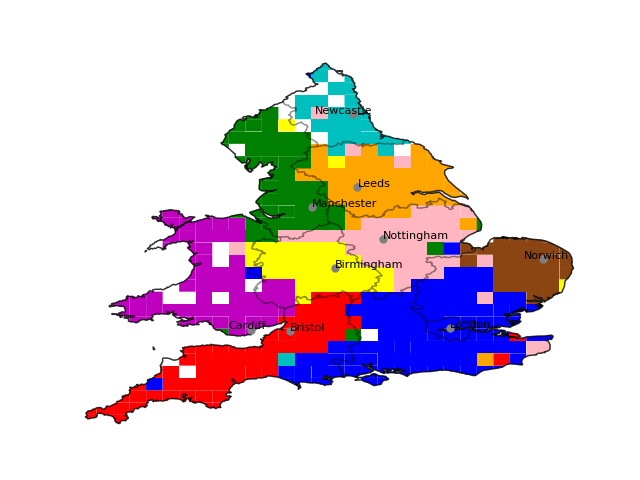}
\caption{Communities in England and Wales determined from Twitter mentions. The Louvain algorithm suggests 9 communities is optimal for this grid resolution. The largest city in each community is labelled. White space means no tweets were recorded. The regions identified correspond roughly with England and Wales' administrative regions (shown in grey on the map), apart from London which subsumes two other administrative regions. }
\label{fig:uk_com}
\end{figure}

Figure \ref{fig:uk_com} shows the resulting regional communities, presented as a spatial grid with each tile coloured by its community label. These communities were found (with modularity $Q=0.209$) using a network constructed using a $30 \times 30$ grid. The sensitivity of community structure on the grid resolution is analysed in Supporting Information 1, finding that communities are quite robust to variation in the size of the grid boxes. Supporting Information 2 shows an image of the network itself, as well as summary statistics.

Examining the map shown in Figure \ref{fig:uk_com}, there is a striking geographical coherence to the communities, with 9 contiguous regions easily identified. There are some `outliers', tiles belonging to a community different than their neighbours. These outliers typically have low populations, and hence low numbers of Twitter users and small edge weights, so their assignment has very little effect on the total modularity.  Overall the communities reflect `folk' preconceptions of where the regions of the UK should be, have a reasonable correspondence to administrative regions, and also agree with previous work using phone call networks \cite{Ratti2010}. The main difference occurs in the London region, where London, the South-East and part of East Anglia are incorporated into one region. This area comprises (broadly) the extent of London's `commuter belt' and is likely an effect of London's enormous economic and cultural influence. Henceforth will we label each region by its largest city, for ease of reference and since, by examination of Figure\ref{fig:compg}, most of the communication volume originates from the largest city in a region. 

\section*{Comparison of Vocabulary}

Given that people are more likely to be `friends' with people living nearby \cite{Backstrom2010}, the types and topics of communications within and between communities may be different. The field of topic modelling, in general and for Twitter, has a large literature (e.g. \cite{Michelson2010,Weng2011,Atefeh2015}). Other research has studied dialect differences on Twitter, particularly in the USA \cite{Huang2016,Blodgett2016,Donoso2017}. Here we use a simple approach to compare the words and topics used in intra- and inter-community communication. 

\begin{figure}[H]
\centering
\includegraphics[width=0.6\textwidth]{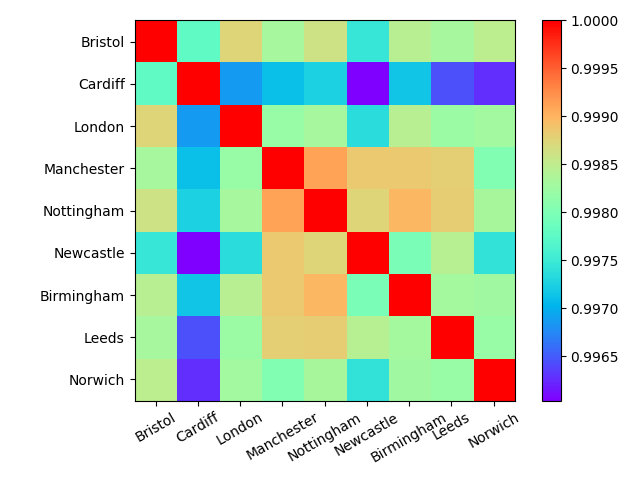}
\caption{Cosine similarity for word vectors corresponding to each community's lexicon. This metric equals 1 for identical vectors. Wales is notably less similar to all other regions while the more northern regions are more similar to each other than the southern ones.}
\label{fig:cosine}
\end{figure}

We first create a lexicon $W$ containing all distinct case-insensitive words ($n=556466$) from all tweets. We removed user names prefaced by an `@' symbol, URLs, special characters (e.g. emojis), as well as the `\#' symbol prefacing hashtags, though we kept the hashtag itself. $W$ defines our word-vector space within which we construct 9 vectors, $\vec{w}_i$, representing the word sets obtained from tweets originating in each of the 9 regions. We use cosine-similarity, $\frac{\vec{w}_i \cdot \vec{w}_j}{\Vert \vec{w}_i \Vert \Vert\vec{w}_j\Vert}$, to measure the similarity between regional vocabulary. Figure \ref{fig:cosine} shows that all regions are quite similar to each other. Cardiff is the most dissimilar, and there is a suggestion that the `northern' regions (Birmingham, Nottingham, Leeds, Manchester and Newcastle) are more similar to each other than to the southern regions (Bristol, London and Norwich). To further investigate this we calculate the tf-idf (term frequency-inverse document frequency) score. Tf-idf assigns high scores to words that differentiate documents within a corpus. We create 9 `documents' by aggregating the tweets originating from each region. The top tf-idf words in Cardiff are mainly local place names or words in the Welsh language. Other regions' highest tf-idf words are mainly local place names or sports clubs (Supporting Information 3). This method successfully detects the Welsh language, while the cosine similarity suggests a small dialect difference between the North and South of England as well as also detecting the difference between Welsh and English tweets.

\subsection*{Local, Regional and National Communication}

To compare how `locals' communicate with each other to how they communicate with `outsiders' we divide tweets originating from a region into two categories: tweets sent within the region and tweets sent to other regions. Let  $f(w)_i^{loc}$ denote word frequencies in tweets sent within community $i$ and $f(w)_i^{out}$ denote word frequencies in tweets sent from community $i$ to any other community. We can then assign each word a frequency rank $r(w)$ from most common ($r(w) = 1$) to least , so $f(w)_i^{loc} \rightarrow r(w)_i^{loc}$ and $f(w)_i^{out} \rightarrow r(w)_i^{out}$. Looking at the rank differences: $\Delta r_i = r(w)_i^{loc} - r(w)_i^{out}$
then gives us a rough indication of how the vocabulary varies in tweets sent within regions compared to those sent between regions. Positive rank differences indicate a word is more common in inter-community messages and negative rank differences indicate the word is more common in intra-community messages. In order to avoid distraction by rare words with spurious large rank differences, we restrict our analyses to words with frequency per tweet (calculated separately in each region and for {\it loc} or {\it out}) greater than 0.1\% for every word considered. We can look at pairs of regions in the same way. We rank the words used to communicate between communities $i$ and $j$, $r(w)_{ij}$, and the words used to communicate between $i$ and all other communities (excluding itself and $j$), $r(w)_{i,k \notin \{i,j\}}$, and look at
$\Delta r_{ij} =r(w)_{i,k \notin \{i,j\}} - r(w)_{ij}$
to see what words are characteristic of communication between $i$ and $j$ specifically. 

\begin{figure}[H]
\centering
\includegraphics[width=0.6\textwidth]{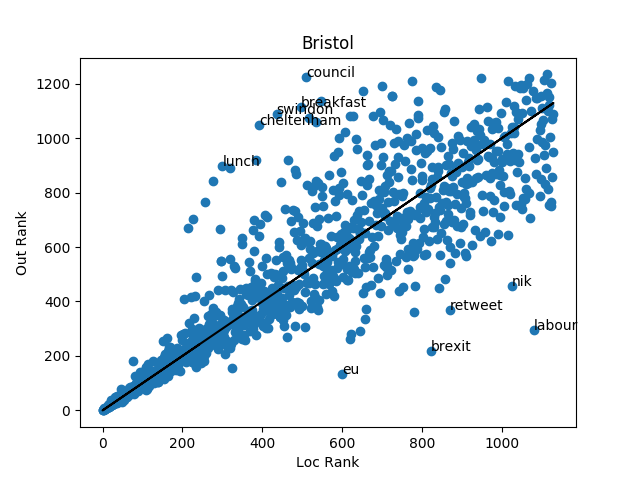}
\caption{Loc rank versus out rank for Bristol region. Words with largest magnitude rank difference are indicated. Intra-region words include local place names while inter-region words refer to national politics. }
\label{fig:word_rank}
\end{figure}

Figure \ref{fig:word_rank} and Table \ref{tab:rankdiff} show that intra-community words (negative rank difference) primarily refer to local issues like sports (pigeonswoop, villa, mufc) and places in the region (wigan, bradford, chester) similar to the high ranking tf-idf words. Inter-community words primarily refer to national issues (brexit, eu, nhs, tory). Table \ref{tab:rankdiff2} shows an example for two neighbouring southern regions. Sport shows up again in pairwise communication, as it does in local communication, but not nationally - indicating that sporting rivalries are playing out on a regional level, as one might expect.

\begin{table}[H]
\centering
%\resizebox{\textwidth}{!}{
\begin{tabular}{|c|c|c|c|}
\hline
London &Manchester &Birmingham &Leeds \\ \hline \hline
liverpool(613) &brexit(694) &brexit(803) &brexit(682) \\ \hline
sleep(369) &government(612) &disabled(663) &eu(594) \\ \hline
trending(337) &tory(572) &extremely(654) &leaving(570) \\ \hline
ff(316) &nhs(506) &inclusion(654) &nhs(539) \\ \hline
topic(304) &eu(493) &wwfc(518) &ref(520) \\ \hline
\hline 
co(-396) &wigan(-475) &thursday(-491) &pop(-497) \\ \hline
brighton(-423) &chester(-504) &pigeonswoop(-504) &south(-504) \\ \hline
lunch(-449) &lunch(-515) &villa(-557) &thursday(-512) \\ \hline
awards(-472) &gt(-564) &art(-692) &council(-564) \\ \hline
greater(-818) &mufc(-606) &blues(-700) &bradford(-586) \\ \hline

\end{tabular}
%}
\caption{Top and bottom five rank differences, $\Delta r_i$, for 4 most populous regions. See Supporting Information 3 for full list.
}\label{tab:rankdiff}
\end{table}

\begin{table}[H]
\centering
\begin{tabular}{|c|c|}
\hline
Rank difference & Cardiff to Bristol \\ \hline
30622 & busygettingbetter \\ \hline
22063 & anglowelshcup \\ \hline
21903 & younglivesvscancer \\ \hline
17430 & bigwednesdayshow \\ \hline
14548 & wenurses \\ \hline
\hline
Rank difference & Bristol to Cardiff \\ \hline
19869 & tywydd \\ \hline
20289 & scarlets \\ \hline
29627 & tandyout \\ \hline
40499 & tandy \\ \hline
40727 & goalscorers \\ \hline
\end{tabular}
\caption{Top and bottom five rank differences, $\Delta r_{ij}$, for Bristol to Cardiff and vice versa. Most of these terms are popular hashtags. Cardiff is talking to Bristol about health related issues, a radio show and a sporting event. Bristol is talking to Cardiff about sports and `tywydd', Welsh for `weather'. }\label{tab:rankdiff2}
\end{table}

\section*{Communication Flow Between Regions}

\begin{figure}[H]
\centering
\includegraphics[width=\textwidth]{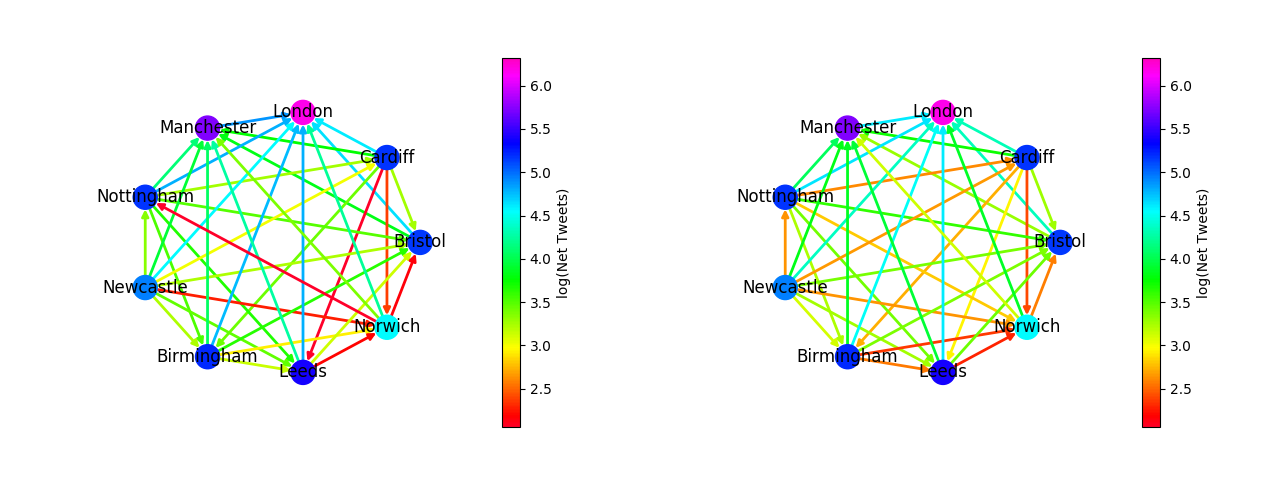}
\caption{Left: Number of mentions sent between UK regions. Arrows show the direction of flow e.g. Manchester mentions London more than London mentions Manchester. Node colour shows number of mentions sent within a community. Right: The flow of mentions computed via the null model, node colours same as left image. }
\label{fig:uk_flow}
\end{figure}

Now that we have an assignment of each tile to a community, based on the undirected network, we form a directed network induced by the assignment of communities in Figure \ref{fig:uk_com}. We want to know about the net flow of mentions i.e. Does London mention Manchester more than vice-versa? Let there be $N$ communities and let $m_{ij}$ be the number of mentions of (users in) community $j$ by (users in) community $i$. If $m_{ij} > m_{ji}$ we draw a directed edge from $i$ to $j$ with weight $m_{ij} - m_{ji}$. If $m_{ij} < m_{ji}$ the edge goes from $j$ to $i$ with weight $m_{ji}-m_{ij}$. Thus our arrows always point towards the region which is mentioned more, weighted by the net difference in communication volume.  We show this network in Figure \ref{fig:uk_flow} (left).

We see that London (the most populous region) is always mentioned more by the other regions than vice versa. This is perhaps expected, as tweets referencing politics are likely to be directed towards the capital. Manchester (containing the second largest city) is mentioned more by all but London. In light of previous work \cite{us} showing that high population density leads to a super-linear increase in the amount of Twitter activity, this is perhaps not surprising; London and Manchester are very densely populated regions, so contain a lot of users and hence present a large `target' for other regions. However population is not a perfect predictor, contrast Newcastle (always in deficit of mentions) with Norwich. Despite Newcastle having a larger population ($\sim$ 2.7 million versus $\sim$ 1.6 million, see Table \ref{tab:ratio}, data from UK Census\footnote{https://www.ons.gov.uk/peoplepopulationandcommunity/populationandmigration/populationestimates (Accessed April 2018)} ) it is mentioned much less than Norwich, perhaps an indication that its geographic isolation (it is far from the large population centres in the North-West and South-East) is leading to some social isolation.

To see which communities talk more or less than expected we establish a null model by cutting all the outgoing edges and rewiring them randomly, while keeping the in- and out-degree of every node fixed, to account for the relative size and activity of each region. This null model generates the expected pattern of inter-region communication based on Twitter activity in each region, assuming no bias in inter-regional communication. We do not redirect self-edges since the communities are, roughly, chosen to maximise self interaction by the Louvain algorithm. Comparison to a null model which randomly reassigns self-edges will thus show that the observed graph has more self interaction, by construction. It is more informative to focus on inter-community communication only. A community $i$ has 
$m_i^{out} = \sum_{j \neq i} m_{ij}$
outgoing mentions and 
$m_i^{in} = \sum_{j \neq i} m_{ji}$
incoming mentions. If mentions are distributed equally to all regions, in proportion to their share of incoming edges, then we would expect the fraction of mentions from $i$ to $j$ to be 
$\bar{m}_{ij} = m_i^{out} \frac{m_j^{in}}{M_i^{in}}$
where $M_i^{in} = \sum_{j \neq i} m_j^{in}$. 

We compare the expected to observed in Figure \ref{fig:uk_flow}. The main observation here is that all regions communicate more with London (and to a lesser degree Manchester) than the null model predicts. The volume of communication between other regions is therefore less than expected, however, the direction of net communication flow is preserved for all pairs but one: Norwich talks more to Nottingham, whereas the null model predicts the opposite.  For each region we look at the ratio of total incoming to total outgoing mentions in Table \ref{tab:ratio}. This paints a similar picture; only London has a ratio greater than one and the  North-East (Newcastle) has the lowest ratio of all 9 regions. 

\begin{table}[H]
\centering
\begin{tabular}{|c|c|c|}
\hline
Region & Population & $\frac{ \sum_{j\neq i} m_j^{in} }{ \sum_{j\neq i} m_j^{out} }$ \\ 
\hline
London & 22650335 & 1.629 \\ \hline
Manchester & 7733168 & 0.981 \\ \hline
Bristol & 4623903 & 0.838 \\ \hline
Leeds & 5571395 & 0.774 \\ \hline
Birmingham & 5038212 & 0.746 \\ \hline
Norwich & 1561239 & 0.733 \\ \hline
Nottingham & 5493704 & 0.690 \\ \hline
Cardiff & 2818706 & 0.689 \\ \hline
Newcastle & 2744728 & 0.638 \\ \hline
\end{tabular}
\caption{Regional populations (using our discovered regions) and ratio of number of incoming mentions to number of outgoing mentions.}\label{tab:ratio}
\end{table}

\section*{Inter-region Sentiment}

Regional identities and rivalries lead to strong emotions about sport, politics or any number of issues. Local stereotypes may lead to negative associations with a particular place. By analysing the text of the messages exchanged between regions we can ask if these expectations are reflected in the sentiment of the communication. 

Sentiment analysis on Twitter is another large topic. Early work used sentiment analysis of tweets to try to predict elections, \cite{Tumasjan2010} movie box-office returns \cite{Thelwall2011} and brand sentiment \cite{Mostafa2013}, demonstrating the power of the approach. Much research has been done on improving sentiment analysis for short texts like tweets or SMS messages \cite{Kiritchenko2014, Eugenio2014, Giachanou2016}. 
We use a popular lexicon-based sentiment analyser \cite{Smedt2012, Loria2014} to to assign a polarity to each tweet. Polarity is a number between -1 and 1 measuring how negative or positive the sentiment is in a text.

\begin{figure}[H]
\centering
\includegraphics[ width=\textwidth]{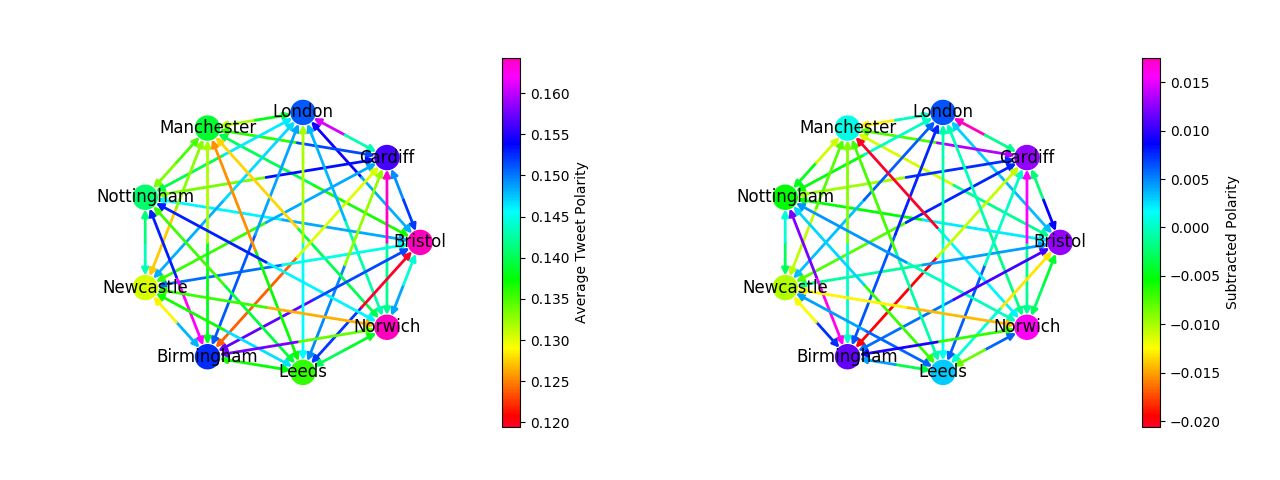}
\caption{Left: Arrows: average sentiment per tweet between regions. Node: average sentiment for tweets sent within region. Right: Arrows $(p_{ij} - \mu_i)$, nodes:  $(p_{ii} - \mu_i)$ i.e. sentiment corrected for baseline of each region. }
\label{fig:uk_sen}
\end{figure}

We explore the message sentiment in two ways, again using the induced graph. Figure \ref{fig:uk_sen} (left) shows the average polarity of a message sent between any two communities, $p_{ij}$. Polarity is on average positive, indicating the average tweet which mentions another user is positive. This is in line with research on sentiment in other corpora which finds a general trend towards positive polarity \cite{Dodds2389}. Self-polarity is shown as the node color, indicating that southern regions are more positive in both inter- and intra-community communication. 

To determine the background level of sentiment for each region, for each community let $\mu_i = \frac{1}{N}\sum_{j} p_{ij}$. Each arrow or node in Figure \ref{fig:uk_sen} (right) shows $\tilde{p}_{ij} = (p_{ij} - \mu_i)$. As differences in regional vocabulary may lead to some regions sending tweets with lower measured polarity scores, this procedure allows us to look at inter-region communication relative to the baseline in each region. See Supporting Information 3 for exact values, with errors. As we have so many mentions, most average polarity measurements are quite precise, smaller than the resolution of the colormap, the largest errors are between distant pairs of small regions e.g. Newcastle and Norwich. Figure \ref{fig:uk_sen} (right) shows that after correcting for background sentiment, southern regions are still relatively more positive about themselves. The `friendliest' pair of regions, i.e. the pair with the highest $\tilde{p}_{ij} + \tilde{p}_{ji}$ are the neighbouring Midland regions Nottingham and Birmingham, while the least friendly are Birmingham and Cardiff. This implies spatial proximity alone does not account for inter-region sentiment.

We calculate two additional metrics based on polarity, $s_i = p_{ii} - \frac{1}{N-1}\sum_{i\neq j} p_{ij}$ and $\tilde{P}_i = \frac{1}{N-1}\sum_{j \neq i} \tilde{p}_{ji}$. $s_i$ measures how positive a region is in communication with itself compared to its communication with other regions, its `self-regard'. $\tilde{P}_i$ measures how positive other regions are about region $i$, its `popularity'. Values are shown in Table \ref{tab:sentiment}. Southern regions have slightly positive $s_i$, so they are more positive about themselves than about other regions, northern regions tend to be neutral or negative.
Perhaps surprisingly, given its centrality in political discussions, London is the region with the highest incoming polarity from the other regions i.e. the most popular.

\begin{table}[H]
\centering
\begin{tabular}{|c|c||c|c|}
\hline
Region & $s_i$ & Region & $\tilde{P}_i$ \\ \hline
Newcastle & -0.0116(8) & Manchester & -0.0108(3) \\ \hline
Nottingham & -0.0059(5) & Newcastle & -0.0049(7) \\ \hline
Manchester & 0.0012(4) & Norwich & -0.0023(8) \\ \hline
Leeds & 0.0036(5) & Leeds & -0.0006(4) \\ \hline
London & 0.0072(2) & Nottingham & -0.0006(5) \\ \hline
Birmingham & 0.0130(5) & Bristol & 0.0014(4) \\ \hline
Cardiff & 0.0145(6) & Birmingham & 0.0038(4) \\ \hline
Bristol & 0.0145(6) & Cardiff & 0.0038(6) \\ \hline
Norwich & 0.0180(11) & London & 0.0042(2) \\ \hline
\end{tabular}
\caption{Self minus outgoing sentiment, $s_i$, and average incoming sentiment,$\tilde{P}_i$, which measures the sentiment of the other regions for the target region. Norwich has the most self-regard, Newcastle the least. London is the most popular region and Manchester is the least.}\label{tab:sentiment}
\end{table}

\section*{Conclusions}

Traditional regional identities are reflected in social media interactions. Located tweets are a unique resource that allow both community identification and analysis of inter-community communication. We have examined the volume of messages sent between regions, the vocabulary/topics  within a region versus the vocabulary/topics used to communicate with other regions, as well as sentiment between regions.

We must be cautious in our analysis and recognise that Twitter (like all surveys, solicited or not) is not giving us an unbiased view of society at large. It is heavily urban \cite{us, Hecht2014} and over-represents e.g. younger, higher-income people \cite{Malik2015}. Twitter is also a platform that is used more to discuss news and social issues than personal communication, in contrast to say LinkedIn or Facebook which have different characteristic uses. This is both a feature, allowing us potential access to contentious or divisive topics, and a bug e.g. in sentiment analysis, we could be examining an unusually negative corpus. This also has consequences for the volume of tweets - the news and politics focus of Twitter is perhaps another reason that London, seat of government, finance, as well as many news organisations, is over-represented. 
%We would also caution against using this as diagnostic of regional stereotypes: e.g. fewer tweets from a region can be due to economic deprivation rather than a taciturn character, this method is silent regarding the reasons for deficits or excesses in tweeting or sentiment.

Nevertheless, this combination of community identification with text analysis has widespread application. Marketing and political campaigns could potentially use this methodology (perhaps at a smaller scale than national) to identify relevant local issues or if they are targeting single or multiple `communities', which may respond better to different messages. Beyond practical applications, this methodology has the potential to build a quantitative, econometric basis for the study of cultural exchange. The agents of this quantitative theory are the emergent regions, and we can use this combination of social-media data, network science and text analysis to shed light on regional discourse, dialect, connectivity or possibly even regional tension in an area more fractious than the UK. This method provides a way to characterise regions and both suggests interesting social questions (e.g. why does Norwich have such a large `influence' relative to its population?) and also provides the empirical data to {\it quantitatively} test explanatory theories. In general we believe this methodology will help expose the relationship between people, social media, space and place. 
\clearpage

\section*{Supporting Information 1}

Here we study how the partition of England and Wales into regions depends on how we aggregate users. 

\subsection*{Varying the Grid Resolution}

We, as in previous work \cite{Ratti2010}, look at the connections between grid tiles rather than users themselves. The size of the grid tile is chosen by us, we simply divide our bounding box into an $X \times X$ grid. Given this somewhat arbitrary choice it is important to investigate the dependence on the chosen grid resolution. 

\begin{figure}[H]
\centering
\includegraphics[height=0.3\textheight, width=\textwidth]{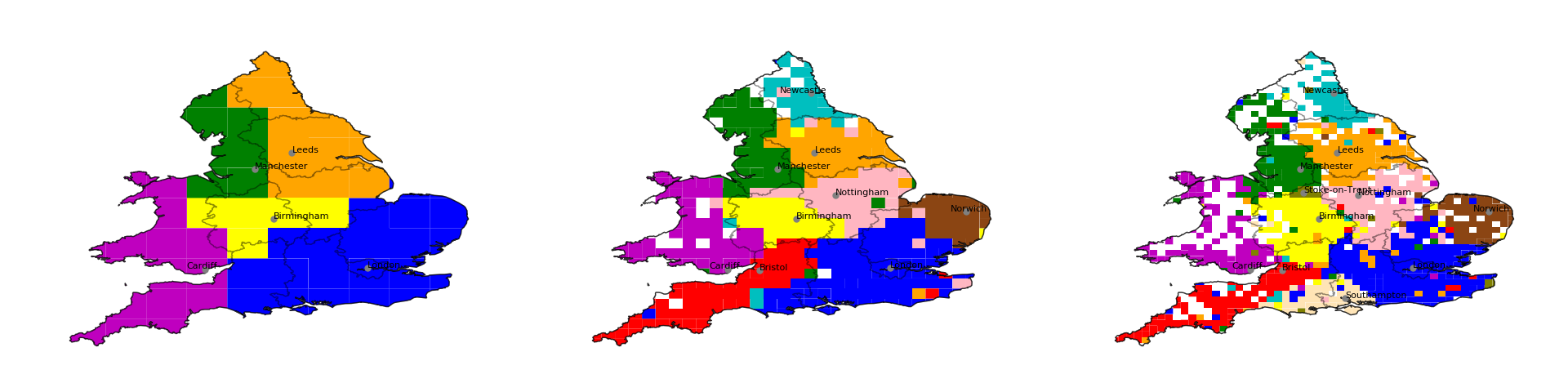}
\caption{Left to right: Coarse, Medium and Fine grids. }\label{fig:compg}
\end{figure}

Figure \ref{fig:compg} shows three grids, Coarse ($10 \times 10$), Medium ($30 \times 30$, which we use in the main text) and Fine ($52 \times 52$). We see that the Coarse grid identifies 5 regions (roughly: South-East, Wales \& South-West, Midlands, North-East and North-West). Clustering on the Medium grid identifies the 9 regions discussed in the text. Clustering on the Fine grid identifies the same 9 regions, plus two small additional regions, centered around the cities of Stoke-on-Trent and Southampton. Modularity increases going from the Coarse (0.086) to Medium (0.209) to Fine (0.255) grid resolutions.

%%% Each figure should be on its own page

There is no {\it a priori} reason to prefer one grid resolution over another. In this case, different choices allow different hierarchical levels of structure to be distinguished. Our choice of the $30 \times 30$ grid for primary study is motivated by the observation that it is the coarsest grid that roughly reproduces the administrative regions of England and Wales. Very small communities have low volumes of communication to/from them and it becomes difficult to make statistically meaningful statements. 

As we increase the resolution further we find further subdivision of the regions e.g. a small part of the South-East splits into its own community centered around Southampton. In the limit where we consider every user individually we find 1000s of communities. Our final choice is a compromise between high modularity and large volumes of communication data flowing between the regions. Furthermore, the same users and tweets are included in the common regions identified by the Medium or Fine grids. This means we would obtain the same results if we performed the analysis using the Fine grid, though with some edge cases differing slightly and some communities splitting at the periphery. Although the modularity is bigger for the Fine grid, these considerations lead us to choose the Medium grid in preference. 
\clearpage

\section*{Supporting Information 2}

We require a non-zero number of internal mentions, i.e. $e_{aa} > 0$, in each tile, this {\it ad hoc} condition removes very sparsely populated tiles which can be assigned to any community without significantly affecting the modularity score. We are left with $N=454$ nodes/tiles in the graph. The associated undirected mention network contained $65934$ edges (i.e. density of $0.641$) with mean node degree $31.4$ and mean weighted node degree $19808.7$. The network has no isolates i.e. it is equal to its giant component. 

%\subsection*{Subhead}

\subsection*{Twitter Mention Network for England and Wales}

%%% Each figure should be on its own page
\begin{figure}[H]
\centering
\includegraphics[height=0.65\textheight]{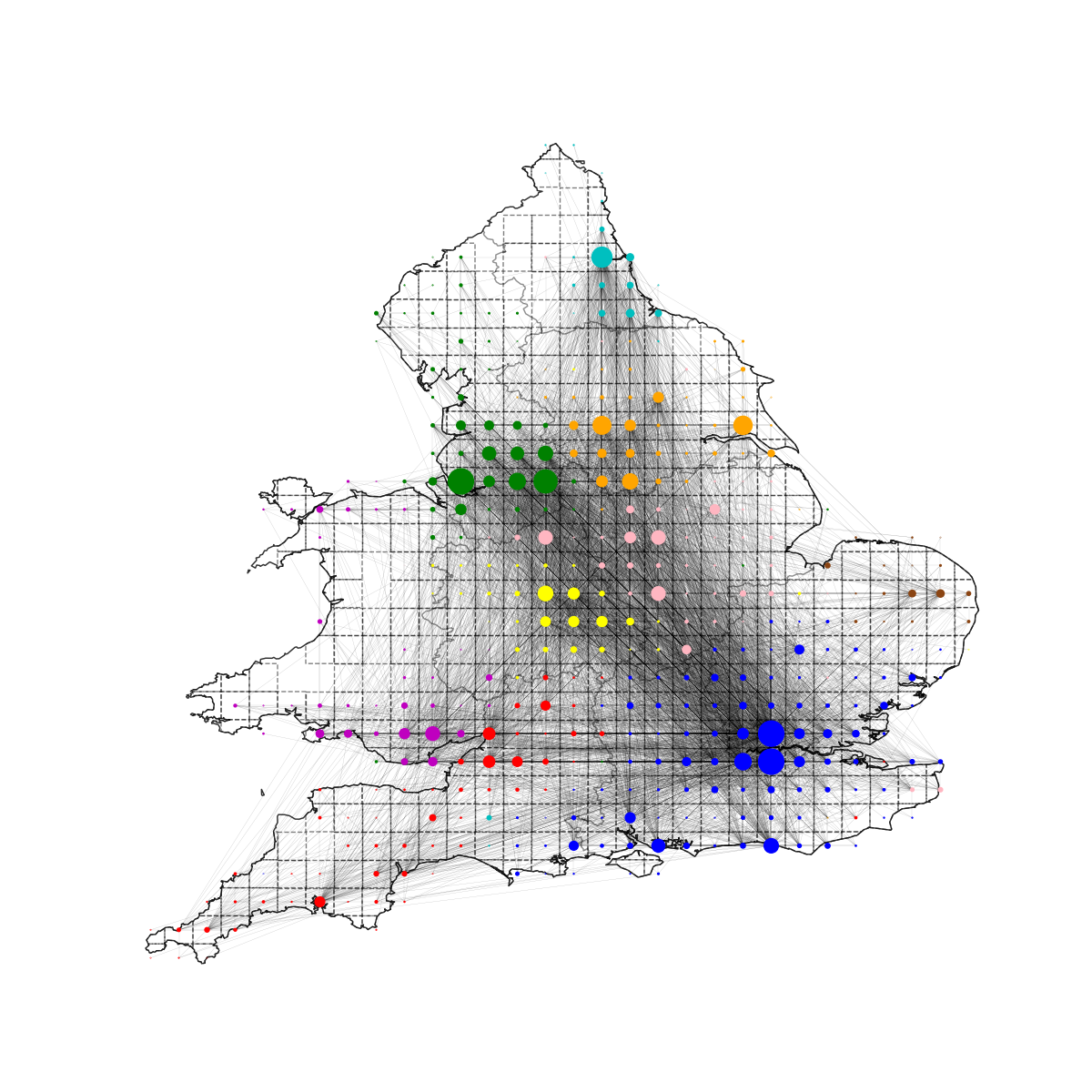}
\caption{The undirected network of Twitter mentions in England and Wales aggregated on a $30 \times 30$ grid. The node sizes correspond to the number of tweets sent within a tile i.e. the size of the self edge $e_{aa}$. Colours correspond to the community allocation discussed in the text. Only connections where $e_{ab} + e_{ba} > 100$ are shown. The map is reflective of population density, showing large numbers of tweets originating in the south-east and north-west. We also see significant communication flow between the regions, supporting our assertion that this data set can be used to study inter-region communication and sentiment. }\label{fig:network}
\end{figure}
\clearpage

\section*{Supporting Information 3}

\subsection*{Top 100 `Local' Terms by TF-IDF.}

{\bf  London  } \newline \newline
{\myfont
trndnl mymyselfandmyphotos afirmremain kenonmkfm greenwichhour ealinghour womened jbombs catford freelancephotographer artconnects drivingforward mcmlvp cosplaygirls countryradio automotiveindustry rbemusicshowcase sneakerjeans mcmcomiccon ldnmcmcomiccon sqsc partyclubmix mcmmidlands willink filteredphotos britishsports forkster noskillphotography cosplayer lewisham controlyourbody southsea farmoor tooting gohartsc wearelondon headington brownout gistwithcynthia commercialvehicle leighonsea cheam hertshour govia iicsa eyesofladyw trumpdossier mylocalculture ukbtsarmy toxicsurv upthehamlet wristinstability cosplaymasquerade lovemk removerutnamskcb southeastern mictizzle ncsupperclubs newham bwsfb ldnmovesme visitbanbury upthestoke muckymerton feedthepoor meaow coycards cobbett cultureofukafrobeats boweldisease wearerugby londonfashionweek thelindaeshow samplesale osteopathy midcarpalinstability dhfc elft runwithandy bostikpremier gamificationeurope ulcerativecolitis northlondonhour otmoor mertonculture southlondon londres lovemycity ifgs herberts digme moveitshow axschat wristpain wandsworth successhour rbwm ruinadatein engagors beckenham } \newline \newline
{\bf  Manchester  } \newline \newline
{\myfont
thefootball oldhamhour marktokerrys beeryview cheerstobeers adelaidies ffrindiau evostikleague lbe leadinggm urmston tweacon tameside camrgb oneport aufc sthelenshour eclcm savemajorcrimes calderstones bolllocks chestertweets corinne itsliverpool cfba mdso fabcan saletown poggy newbrew stav kirbs larklane cumbrianbeer isplenty churchgate theys kerrys prestonhour ourmanchester bevys faithinyoungpeople ancoats ministryofsport knitswithbeer jukeboxthursdays manchestercentralfc swanfest gmcuk purpleandproud janeballlandscapes wxmafc codarmy gan prosandcoms pieceofrubbish oafc loveforever upthetics prelovedhour chorlton majorcrimes lancashirehour sthelens wallasey fmcreators thinkfaster wythenshawe bestshowontv teamukfast onesalefc timperley proudtofoster proudtoadopt uhn mancmade teamorrell janeballphotography raggies wwoolfall altrincham indiehour mcrchristmas ourtownsteam rra hulme piccadillyward lunchtimers dazzzzer chorley cwfl moston poggys hogblog slingthemesh pnefc gbar fitnessisakeyrequirement baltictriangle refperformance } \newline \newline
{\bf  Bristol  } \newline \newline
{\myfont
niks cornwallgate exeterhour bathindiechat swindonhour ukcachehour cmonbris mojorocks communityownership somersethour cornishhake exeterlive freshcornish countrykids devourbrischat champrugby nerdschatting remoanersbs mikevigorfanclub ourchelt dailyinvinciblecover wnukrt theendofallthings glawsfamily devonhour englishriviera makingbristolproud savebathfromboring crediton daretodisrupt selworthy smwbristol musichouruk philskillerthree glosbiz barnstaple ericalovesyou packthegate torbayhour winchcombe torbay merrybrexmas lovethebarbican bristolrugby womaninbiz teahour jacki officialgettogether newlyn walt nsomerset exmoor finzelsreach tiverton cind glaws hagx bewhatyouare cockington swtawards cornishfootball ukwinehour signedupsaturday remainfakenews bris futurecity saltash babbacombe capetown devonfoodhour wickedontour nbtproud babber opmfamily amazingarchie cwlbirds treesofthisworld paignton teignmouth glos boosttorbay brivbed maymedia cornishlove remainlies veawards hetourism bristolandproud truro gloshour falmouth poyf propertyladder newtonabbothour eatmorehake transhealth amateurradio bristolbizhour topsham dartmoor } \newline \newline
{\bf  Leeds  } \newline \newline
{\myfont
euers yesladhull kirklate sheffieldissuper hcafc wharfedalelinks ukcolumn safclive htafc kirklatemarket bcafc euer gmy gewgo doncasterisgreat lovinleeds bigupbradford weareyork cookridge dearne connectingthepieces northstarchat hullhour theworldiswatching maboxing headingley realaleawaydays tvaddict dts ofosheffield hazlegreaves utm marketingweek standwithduchess ukwinehour backinghuddersfield sheffbeerweek ehab calumshullcrew topoftown fanthursday getskyhome nyp leedshour yesladlogo brexitnow syp dazl fido teamdylan otweeklds ijw gayleeds barnsleyisbrill nomorefelling filey taddy teamtheo scc allam projekt horsforth nostell halfhourofpower torbayhour gafc isdead stophs amey geofleurshop bdxmas dewsbury saveshefftrees teambentsgreen dwba wahaw ramember selloutsaturday ataw ryuvoelkel liff teambradford meersbrook sheffieldhour allams geddes hkr burnsy goc menston calderdale roundhay cits fearthefin midstaff supportlocalmarkets enjoyyourprivilege abilitiesnotdisabilities teamyas brighouse } \newline \newline
{\bf  Cardiff  } \newline \newline
{\myfont
yn introbiz dwi oneclubonecounty introbizexpo herefordgoals usw ddim wedi jdwpl valueadded betws iawn nawr roath edrych llongyfarchiadau ukdiscoversaudi chdi ymlaen efo oedd alfieandwilf heddiw sportcardiff mewn ydi hynny rbcdn jofli heno redbalwncoch ytad letsgodevils yna hefyd cael llanelli ond gyfer savemajorcrimes mwynhau uswrugby uptheport daysofselfcare bawb ffordd theglovesareon meddwl caerdydd murenger fel wruin wrth aberdare pubofdreams ndwg wneud pigparker heretoplay siarad penarth gyda gyd poorparkingcdf hellocolin hefo neud bigtuesdayshow uhw erbyn gymraeg rwan fynd ringland ysgol depressionrecovery llanedeyrn rhondda rhaglen cardiffmet llanishen newydd bethespark chwarae angen vitallis llandaff ystrad wyburn wearredforwalesandvelindre cathays sydd virtualworldrideforrity ammanvalley sesiwn siwr grangetown gweld roedd } \newline \newline
{\bf  Norwich  } \newline \newline
{\myfont
norfolkhour skullswasps livinghistory allezallezallez norwichhour euroland theveryear suffolkhour upthepeckers holtbirdcafe norfolkfootball felixstowe alycia yarmouth uea tryyy autisticandproud autismaware norvwas wickeduk charitygala bigdaddyfamily badtouch rolloff facoachmentor liveworklovenorfolk garboldisham burystedmunds wearewsft drinkextraordinary northnorfolkcoast fidel caw marketmen cromer vhurrell oldchimneysbrewery nmfc erwfl holkham soyuztour akingsransom planetsuffolk weareabode wasvlei tryyyy autisticsinger promnation proudcanariesfc edpphotographer lavenham pltv forecastingchallenge grandmentors pdnreflect canarycall ueailm lovethekick communitypub gomagpies dereham beccles mfta swanlavenham commitioner srbeny norwichpubs wasvhar klose pricklypals greatyarmouth teamnnuh fawomenscup itlfc pricklypal tayfen cihcareersweek otbc fawpl woodbridge hotelschool teamwicked emwfl kyley plasterersarms sudbury thetford harvwas needhammarketfc kingslynn stowmarket hoglet sundayatthemusicals holthogcafe trowse norfolkrestaurantweek buildthenest norwichlighttunnel norwichlights foxythanks } \newline \newline
{\bf  Birmingham  } \newline \newline
{\myfont
covhour morphettes thisiscoventry brumhour worcestershirehour jewelleryquarter covhourlive leamingtonhour sociallyshared noflyzone kingsheath xhx lovebrum suahour bizw ipex cwrocks bilbrook thepaguide lovelydrop eastsidejazzclub nearbywild digbeth tuffnut pusb kfe progmill sutcolhour pedsicu thekindnessofpeople salop newbuckshead harrisgibbshair dayswild jonworcesterman willhomewoodbirmingham afrin soulstew secondsinbrum warwickshire tonetaxi covcab grumpys teamwolves cwbf atruegent loveleam harborne biggen gbccexpo nowplaying yeworcestershire blazemotm rockly yecompany studley wearebeingwarned paintwithcars mtptproject haircolour silvermountainagency dosanjh carlingo wehavebeenwarned brumbloggers wolvesaywe stirchley bypy moleend lovesolihull chamberlinkdaily bcccawards hellobrum whitneyout nuno tasteofsrilanka olton pyfproms ckob blackcountry malvernhills jq birminghamartists bonser bubblebobauk lifeecho fosun bridgnorth huis mijn zyderney getinvolvedbrum thriveawards glassboys pelsall brumindependents brumtechawards fwaw boldmere malvernhillshour } \newline \newline
{\bf  Nottingham  } \newline \newline
{\myfont
lollol nccc wherehistorybegins fankew blatherwick nonlge doigy dilnot pineappleoregg whowzers dogtanian lbororugby cousion betteridge cousions jaqui luking gudvkuking boppy simmos bforbhour gasteric byepass changingbeers jannet teamdmu blatherwicks proo earlycrew panthersnation tkofficialmc sethslegacy lborowalksonwater doigys hauntedthursday askuon psyed bennifit includeing footbsllers rebete agentr disolsys lborofamily thff picheal teambeswicks nantwich lboro earings pvfc trentham westerne lovestoke pancanstory lwow southerton wearelincoln votecomedy jsnnet stokeontrent faydeesupporter greennwhites wlechat teamngh armyred boppys farndale proudtobestaffs thrapston nffcclub faceofsot bfbd farndales twitterposh ripleys faydeearmy hcfc antoney greenandgold dilnots djfestlei nearley marksfacuptour loveintheafternoon togetherforautism leicscomedyfest myk globelist tennereefe valproate fatrophyquest theverseandguests lanzorottee ukmodel jasondenoshow kenty laughterloft welovenewhomes fingertoescrossed } \newline \newline
{\bf  Newcastle  } \newline \newline
{\myfont
votecategories consett nefollowers tnarmy prayfortidal leadgate voterfraud coymmp pelaw sparkysrunningclub sundayya totalsport votingcountmodel cheryll northeasthour morekidsonbikes enlscores goffy mancrushforever thisismine northeast corbynistalife northeastcreatives newcastlescaleupsummit letsgoeagles newcastlegateshead veteranslivesmatter utb howayblyth stillflowering hawaythescholars cillamoment templeofboom morpeth getnorth epdp uptheglebe digitalshowcase ultravires boldon hebburn metrocentre cdsou teesvalley teesside chesterlestreet stopthehate welcometosunderland southshields allhailtotheale tidesoflove turfeoke justwrestle amandain madeinnewcastle thisnortherngirlcan edinburghisblackandwhite gateshead hitthebar shinethelightbrightaootherscansee ouseburn aycliffe cramlington tvbcmembers robmcavoy derwentside sackrodwell narrowmind heworth stockton mackem votercountmodel seaham whitleybay uofefutsal cakelicious nonpolitical intothewoods nebloggers boostyourbusiness blobbys babyitschristmas trfc kone bewateraware jesmond teammcjonesforthewin sunderlandfutures boty secundus uptheesh mushroomworks weekendanthems nulli inrafawetrust longbenton roker teammcjones newcastleupontyne haway } \newline \newline

\subsection*{Largest Magnitude Rank Difference Words}

\begin{table}[H]
\centering
\resizebox{\textwidth}{!}{
\begin{tabular}{|c|c|c|c|c|c|c|c|c|}
\hline
London &Manchester &Bristol &Leeds &Cardiff &Norwich &Birmingham &Nottingham &Newcastle \\ \hline
\hline
liverpool(613) &brexit(694) &labour(783) &brexit(682) &brexit(709) &song(597) &brexit(803) &eu(787) &country(567) \\ \hline
sleep(369) &government(612) &brexit(603) &eu(594) &eu(653) &hell(533) &disabled(663) &brexit(696) &mum(558) \\ \hline
trending(337) &tory(572) &nik(568) &leaving(570) &series(580) &cannot(507) &extremely(654) &labour(568) &bro(548) \\ \hline
ff(316) &nhs(506) &retweet(500) &nhs(539) &nhs(568) &film(481) &inclusion(654) &nhs(442) &vote(538) \\ \hline
topic(304) &eu(493) &eu(467) &ref(520) &labour(537) &answer(461) &wwfc(518) &tour(437) &sweet(508) \\ \hline
\hline
co(-396) &wigan(-475) &lunch(-596) &pop(-497) &ar(-564) &students(-515) &thursday(-491) &dave(-509) &nowt(-502) \\ \hline
brighton(-423) &chester(-504) &breakfast(-617) &south(-504) &coach(-570) &latest(-517) &pigeonswoop(-504) &lincoln(-520) &boro(-585) \\ \hline
lunch(-449) &lunch(-515) &swindon(-654) &thursday(-512) &photos(-582) &involved(-522) &villa(-557) &steve(-530) &students(-656) \\ \hline
awards(-472) &gt(-564) &cheltenham(-656) &council(-564) &project(-597) &ncfc(-529) &art(-692) &council(-542) &gateshead(-677) \\ \hline
greater(-818) &mufc(-606) &council(-716) &bradford(-586) &iawn(-852) &lunch(-664) &blues(-700) &aswell(-881) &safc(-778) \\ \hline
\end{tabular}
}
\caption{Top and bottom five rank differences for all regions. 
}\label{tab:rankdiffall}
\end{table}

\section*{Inter-Region Sentiment}

\begin{table}[H]
\centering
\resizebox{0.25\textwidth}{0.1\textheight}{
\begin{tabular}{|c|c|c|}
\hline
Region $i$ & Region $j$ & $p_{ij}$ \\ \hline
Bristol & Bristol & 0.164(1) \\ \hline
Bristol & Cardiff & 0.149(2) \\ \hline
Bristol & London & 0.154(1) \\ \hline
Bristol & Manchester & 0.140(1) \\ \hline
Bristol & Nottingham & 0.146(2) \\ \hline
Bristol & Newcastle & 0.150(3) \\ \hline
Bristol & Birmingham & 0.157(2) \\ \hline
Bristol & Leeds & 0.152(2) \\ \hline
Bristol & Norwich & 0.149(4) \\ \hline
\end{tabular} } \qquad
\resizebox{0.25\textwidth}{0.1\textheight}{
\begin{tabular}{|c|c|c|}
\hline
Region $i$ & Region $j$ & $p_{ij}$ \\ \hline
Cardiff & Bristol & 0.152(2) \\ \hline
Cardiff & Cardiff & 0.156(1) \\ \hline
Cardiff & London & 0.161(1) \\ \hline
Cardiff & Manchester & 0.136(1) \\ \hline
Cardiff & Nottingham & 0.134(3) \\ \hline
Cardiff & Newcastle & 0.136(5) \\ \hline
Cardiff & Birmingham & 0.124(2) \\ \hline
Cardiff & Leeds & 0.150(3) \\ \hline
Cardiff & Norwich & 0.142(6) \\ \hline
\end{tabular} } \qquad
\resizebox{0.25\textwidth}{0.1\textheight}{
\begin{tabular}{|c|c|c|}
\hline
Region $i$ & Region $j$ & $p_{ij}$ \\ \hline
London & Bristol & 0.148(1) \\ \hline
London & Cardiff & 0.143(1) \\ \hline
London & London & 0.151(0) \\ \hline
London & Manchester & 0.132(1) \\ \hline
London & Nottingham & 0.140(1) \\ \hline
London & Newcastle & 0.148(1) \\ \hline
London & Birmingham & 0.151(1) \\ \hline
London & Leeds & 0.146(1) \\ \hline
London & Norwich & 0.143(2) \\ \hline
\end{tabular} } \\ \vspace{1cm}
\resizebox{0.25\textwidth}{0.1\textheight}{
\begin{tabular}{|c|c|c|}
\hline
Region $i$ & Region $j$ & $p_{ij}$ \\ \hline
Manchester & Bristol & 0.137(1) \\ \hline
Manchester & Cardiff & 0.152(2) \\ \hline
Manchester & London & 0.138(0) \\ \hline
Manchester & Manchester & 0.139(0) \\ \hline
Manchester & Nottingham & 0.133(2) \\ \hline
Manchester & Newcastle & 0.127(2) \\ \hline
Manchester & Birmingham & 0.138(1) \\ \hline
Manchester & Leeds & 0.137(1) \\ \hline
Manchester & Norwich & 0.140(3) \\ \hline
\end{tabular} } \qquad
\resizebox{0.25\textwidth}{0.1\textheight}{
\begin{tabular}{|c|c|c|}
\hline
Region $i$ & Region $j$ & $p_{ij}$ \\ \hline
Nottingham & Bristol & 0.148(2) \\ \hline
Nottingham & Cardiff & 0.154(3) \\ \hline
Nottingham & London & 0.146(1) \\ \hline
Nottingham & Manchester & 0.135(1) \\ \hline
Nottingham & Nottingham & 0.141(1) \\ \hline
Nottingham & Newcastle & 0.143(3) \\ \hline
Nottingham & Birmingham & 0.162(1) \\ \hline
Nottingham & Leeds & 0.139(1) \\ \hline
Nottingham & Norwich & 0.146(4) \\ \hline
\end{tabular} } \qquad
\resizebox{0.25\textwidth}{0.1\textheight}{
\begin{tabular}{|c|c|c|}
\hline
Region $i$ & Region $j$ & $p_{ij}$ \\ \hline
Newcastle & Bristol & 0.145(3) \\ \hline
Newcastle & Cardiff & 0.148(4) \\ \hline
Newcastle & London & 0.147(1) \\ \hline
Newcastle & Manchester & 0.133(2) \\ \hline
Newcastle & Nottingham & 0.141(3) \\ \hline
Newcastle & Newcastle & 0.130(1) \\ \hline
Newcastle & Birmingham & 0.148(3) \\ \hline
Newcastle & Leeds & 0.147(2) \\ \hline
Newcastle & Norwich & 0.126(7) \\ \hline
\end{tabular} } \\ \vspace{1cm}
\resizebox{0.25\textwidth}{0.1\textheight}{
\begin{tabular}{|c|c|c|}
\hline
Region $i$ & Region $j$ & $p_{ij}$ \\ \hline
Birmingham & Bristol & 0.152(2) \\ \hline
Birmingham & Cardiff & 0.130(2) \\ \hline
Birmingham & London & 0.149(1) \\ \hline
Birmingham & Manchester & 0.132(1) \\ \hline
Birmingham & Nottingham & 0.153(2) \\ \hline
Birmingham & Newcastle & 0.129(3) \\ \hline
Birmingham & Birmingham & 0.152(1) \\ \hline
Birmingham & Leeds & 0.138(2) \\ \hline
Birmingham & Norwich & 0.134(4) \\ \hline
\end{tabular} } \qquad
\resizebox{0.25\textwidth}{0.1\textheight}{
\begin{tabular}{|c|c|c|}
\hline
Region $i$ & Region $j$ & $p_{ij}$ \\ \hline
Leeds & Bristol & 0.119(2) \\ \hline
Leeds & Cardiff & 0.133(3) \\ \hline
Leeds & London & 0.132(1) \\ \hline
Leeds & Manchester & 0.125(1) \\ \hline
Leeds & Nottingham & 0.135(2) \\ \hline
Leeds & Newcastle & 0.137(2) \\ \hline
Leeds & Birmingham & 0.137(2) \\ \hline
Leeds & Leeds & 0.136(1) \\ \hline
Leeds & Norwich & 0.139(4) \\ \hline
\end{tabular} } \qquad
\resizebox{0.25\textwidth}{0.1\textheight}{
\begin{tabular}{|c|c|c|}
\hline
Region $i$ & Region $j$ & $p_{ij}$ \\ \hline
Norwich & Bristol & 0.144(4) \\ \hline
Norwich & Cardiff & 0.164(6) \\ \hline
Norwich & London & 0.148(1) \\ \hline
Norwich & Manchester & 0.128(3) \\ \hline
Norwich & Nottingham & 0.153(5) \\ \hline
Norwich & Newcastle & 0.136(7) \\ \hline
Norwich & Birmingham & 0.158(4) \\ \hline
Norwich & Leeds & 0.140(4) \\ \hline
Norwich & Norwich & 0.164(1) \\ \hline
\end{tabular} } \vspace{0.25cm}
\caption{Average polarity of mentions sent from region $i$ to region $j$. Number in bracket is error on last digit. 
}\label{tab:suppsentiment}
\end{table}

\clearpage

\nolinenumbers

% Either type in your references using
% \begin{thebibliography}{}
% Text
% \end{thebibliography}
%
% or
%
% Compile your BiBTeX database using our plos2015.bst
% style file and paste the contents of your .bbl file
% here. See http://journals.plos.org/plosone/s/latex for 

\end{document}